\documentclass{easychair}

\usepackage{doc}
\usepackage{mathtools,stmaryrd,amssymb,amsfonts,nicefrac,bm}
\usepackage{subcaption}
\usepackage{hyperref}
\usepackage[capitalise,noabbrev]{cleveref}
\usepackage{graphicx}
\usepackage{multirow}
\usepackage{proof}
\usepackage{pgfplots}
\usepackage{pgfplotstable}
\pgfplotsset{compat=1.18}
\usepackage{soul}
\usepackage{todonotes}
\usepackage{cite}

\newtheorem{problem}{Problem}
\newtheorem{example}{Example}
\newtheorem*{remark}{Remark}

\newtagform{Enc1}{(}{)$_{e1}$}
\newtagform{Enc2}{(}{)$_{e2}$}

\DeclareMathOperator{\id}{\mathsf{id}}
\DeclareMathOperator{\gen}{\mathsf{gen}}
\DeclareMathOperator{\swap}{\mathsf{swap}}
\DeclareMathOperator{\seq}{\mathsf{seq}}
\DeclareMathOperator{\stack}{\mathsf{par}}

\DeclareMathOperator{\win}{\mathsf{dom}}
\DeclareMathOperator{\wout}{\mathsf{cod}}

\usepackage{tikz}
\usetikzlibrary{arrows.meta, positioning}
\usepackage{tikzit}
\tikzstyle{gate}=[fill=white, draw=black, shape=rectangle, minimum height=0.43cm, minimum width=0.43cm, inner sep=0.1em]
\tikzstyle{control}=[fill=black, draw=black, shape=circle, scale=0.38]
\tikzstyle{not}=[shape=circle, path picture={ 
\draw[black](path picture bounding box.north) -- (path picture bounding box.south) (path picture bounding box.west) -- (path picture bounding box.east);
}, draw=black, scale=.8]
\tikzstyle{wcontrol}=[fill=white, draw=black, shape=circle, scale=0.38]
\tikzstyle{bwcontrol}=[draw=black, shape=circle, scale=0.38, path picture={
        \fill[white] (path picture bounding box.center) circle(0.5);
        \fill[black] (path picture bounding box.center) [radius=0.5] -- ++(0:0.5) arc[start angle=0, end angle=180, radius=0.5] -- cycle;
    }, rotate=45]
\tikzstyle{empty}=[fill=white, draw=black, shape=rectangle, inner sep=0.4em, emptyborder]
\tikzstyle{globalphase}=[fill=white, draw=black, inner sep=0.15em, shape=rounded rectangle, minimum height=0.4cm]
\tikzstyle{ancilla}=[fill=black, draw=black, shape=rectangle, minimum width=0.01cm, minimum height=0.25cm, inner sep=0.01em]
\tikzstyle{ground}=[fill=white, path picture={\draw[black](-1.5mm,0)--(-0.6mm,0);\draw[black,thick](-0.6mm,-1.75mm)--(-0.6mm,1.75mm) (0mm,-0.9mm)--(0mm,0.9mm) (0.6mm,-0.5mm)--(0.6mm,0.5mm);}, minimum width=0.1mm, draw=none, outer sep=0pt]
\tikzstyle{gate22}=[fill={rgb,255: red,220; green,220; blue,220}, draw=black, shape=rectangle, minimum height=.68cm, minimum width=0.6cm]
\tikzstyle{void}=[shape=rectangle, minimum height=0.5cm]
\tikzstyle{gate44}=[fill={rgb,255: red,220; green,220; blue,220}, draw=black, shape=rectangle, minimum height=1.43cm, minimum width=0.5cm]
\tikzstyle{divider}=[fill={rgb,255: red,220; green,220; blue,220}, draw=black, shape border rotate=90, regular polygon, regular polygon sides=3, inner sep=1.5pt, rounded corners=0.5mm]
\tikzstyle{gatherer}=[fill={rgb,255: red,220; green,220; blue,220}, draw=black, shape border rotate=-90, regular polygon, regular polygon sides=3, inner sep=1.5pt, rounded corners=0.5mm]
\tikzstyle{hyperedge}=[fill=white, draw=black, shape=rectangle, rounded corners=0.1cm, minimum height=.6cm, minimum width=.6cm]
\tikzstyle{square}=[fill=white, draw=black, shape=rectangle, minimum height=0.20cm, minimum width=0.20cm, inner sep=0.1em, thick]
\tikzstyle{gphase}=[rounded rectangle, rounded rectangle arc length=120, fill={zx_grey}, inner sep=2pt, font={\tiny\boldmath}, label distance=1mm, fill opacity=.8, text opacity=1, tikzit category=ZX]
\tikzstyle{customcontrol}=[fill=white, draw=black, inner sep=0.1em, shape=rounded rectangle, minimum height=0.2cm]
\tikzstyle{whiteancilla}=[fill=white, draw=black, shape border rotate=-90, regular polygon, regular polygon sides=3, inner sep=1.5pt, rounded corners=0.2mm]
\tikzstyle{blackancilla}=[fill=black, draw=black, shape border rotate=-90, regular polygon, regular polygon sides=3, inner sep=1.5pt, rounded corners=0.2mm]
\tikzstyle{greyancilla}=[fill={rgb,255: red,150; green,150; blue,150}, draw=black, shape border rotate=-90, regular polygon, regular polygon sides=3, inner sep=1.5pt, rounded corners=0.2mm]
\tikzstyle{whiteancillaterm}=[fill=white, draw=black, shape border rotate=90, regular polygon, regular polygon sides=3, inner sep=1.5pt, rounded corners=0.2mm]
\tikzstyle{blackancillaterm}=[fill=black, draw=black, shape border rotate=90, regular polygon, regular polygon sides=3, inner sep=1.5pt, rounded corners=0.2mm]

\tikzstyle{emptyborder}=[-, dash pattern=on 0.16em off 0.16em on 0.16em off 0.16em on 0.16em off 0em]
\tikzstyle{etc}=[-, draw=black, densely dashed, thick]
\tikzstyle{greyetc}=[-, draw={rgb,255: red,161; green,161; blue,161}, densely dashed, thick]
\tikzstyle{dots}=[-, dotted, draw=black, thick]
\tikzstyle{big}=[-, thick]
\tikzstyle{register}=[-, double]
\tikzstyle{grey}=[-, draw={rgb,255: red,161; green,161; blue,161}]
\tikzstyle{border}=[-, fill=white]

\input{qc.tikzdefs}
\newcommand{\tf}[1]{\scalebox{0.77}{\input{#1.tikz}}}

\newcommand{\smalltf}[1]{\scalebox{0.72}{\input{#1.tikz}}}

\newcommand{\N}{\mathbb{N}}

\newcommand{\defeq}{\coloneqq}

\newcommand{\semicolon}{\fatsemi}

\newcommand{\cat}[1]{\textbf{\textup{#1}}}

\newcommand{\ie}{\text{i.e.,}}
\newcommand{\vampire}{\textsf{Vampire}}
\newcommand{\cvc}{\textsf{cvc5}}
\newcommand{\tptp}{\textsf{TPTP}}
\newcommand{\smt}{\textsf{SMT-LIB}}
\newcommand{\isabelle}{\textsf{Isabelle}}
\newcommand{\lp}{\textsf{LambdaPi}}
\newcommand{\alethe}{\textsf{Alethe}}
\newcommand{\eunoia}{\textsf{Eunoia}}
\newcommand{\lsfc}{\textsf{LSFC}}
\newcommand{\eprover}{\textsf{E}}
\newcommand{\twee}{\textsf{Twee}}
\newcommand{\egg}{\textsf{egglog}}
\newcommand{\zthree}{\textsf{Z3}}

\usepackage{xcolor}

\title{Challenging Benchmarks for Diagrammatic Equivalence of Circuits in \tptp{} and \smt{}}
\titlerunning{Challenging Benchmarks for Diagrammatic Equivalence of Circuits}
\author{Julie Cailler\inst{1}\and Noé Delorme\inst{1}\and Sophie Tourret\inst{1,2}}
\authorrunning{J.\ Cailler, N.\ Delorme and S.\ Tourret}
\institute{University of Lorraine, CNRS, INRIA, LORIA, Nancy, France \and
Max Planck Institute for Informatics, Saarbrücken, Germany}

\begin{document}

\maketitle

\begin{abstract}
     We introduce a new family of benchmarks for the problem of diagrammatic equivalence between circuits. Three variants of this problem are considered, ranging from basic to challenging, and benchmarks are generated for each variant. We provide first-order encodings in both \tptp{} and \smt{} formats, together with scripts that automatically generate benchmark instances, and evaluate these benchmarks on state-of-the-art automated theorem provers and SMT solvers.
\end{abstract}

\makeatother

\section{Introduction}
\label{sec:intro}
A recurrent complaint from benchmark repository maintainers and competition organizers is the lack of new contributions to their repositories.\footnote{It is discussions on this very topic at the Dagstuhl seminar 25441 that motivated the existence of this paper.} It is already a time-consuming task to prepare an archive for reproducibility on an open repository such as Zenodo \cite{zenodo}. 
The extra effort of curating and documenting a selection of benchmarks for addition in a dedicated repository such as \smt{} or the \tptp{} library is important to the research community, but not well rewarded and thus not often undertaken. This paper goes against that trend. It describes a family of benchmarks for first-order automated theorem provers (ATPs) with arithmetic, as well as the tools to use to produce more of them.

The benchmarks and tools presented in this paper originate from an ongoing effort to produce certificates for proofs of diagrammatic equivalence between quantum circuits \cite{trs}. 
In recent years, graphical languages for quantum circuits have been extensively studied, leading to complete equational theories that enable circuits to be manipulated and rewritten while preserving their semantics \cite{completeness,extensions,minimality,controlledprop}. More broadly, quantum circuits are instances of the well-established framework of diagrammatic reasoning, which represents processes as string diagrams and captures their interactions in space and time through sequential and parallel composition \cite{Selinger2011}. These developments open the way to automated tasks such as circuit optimization, simplification, and formal verification.

In this paper, string diagrams are often referred to as \emph{circuits}. Roughly speaking, circuits consist of generators---corresponding to elementary primitives---depicted as boxes connected by wires, and there are many diagrammatic representations of a given evolution. For instance, the two following circuits are equivalent.
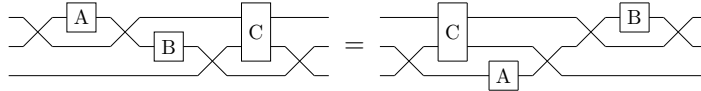
\begin{gather*}
    \tf{example_left}=\tf{example_right}
\end{gather*}

Formally, circuits are expressed as terms freely generated from a set of generators under sequential and parallel composition. As a consequence, syntactically different terms may express the same circuit, depending on the order and structure in which compositions are formed. Therefore, the intuitive notion of diagrammatic equivalence is fully captured by a collection of coherence equations \cite{Selinger2011}, which can be used to rewrite equivalent terms. This naturally leads to the problem of diagrammatic equivalence: deciding, given two terms, whether they can be rewritten into one another using the coherence equations.

We provide tools for generating diagrammatically equivalent terms, which produced the benchmarks studied in this work. In the context of our original motivation---namely, the certification of quantum circuit equivalences---the modest performance of ATPs on these benchmarks, together with the lack of maturity of the associated certificate production techniques, highlights both the difficulty of the problem and the need for further progress
in arithmetic and equational reasoning. For these reasons, we believe that the benchmarks themselves are of independent interest to the ATP community, as a challenge.

We consider three variants of the diagrammatic equivalence problem. The first one is the most general and corresponds exactly to the settings described above. In the second one, we consider circuits with no generators (i.e., no boxes) that allow the expression of any permutation. Finally, the last one is a restriction of the second one to a simpler configuration.

In this paper, we describe the three variants of the problems (\cref{sec:pb_desc}) and explain the encoding built for them (\cref{sec:encodings}). We also discuss the tools used to create the benchmarks (\cref{sec:generators}) and the evaluation of these benchmarks on state-of-the-art ATPs (\cref{sec:eval}). The tools and benchmarks used to run the experiments are available on Zenodo~\cite{zenodo_scripts,zenodo_benchs}.

\section{Problem Description}
\label{sec:pb_desc}
Diagrammatic reasoning and its underlying notion of \emph{circuits} is captured formally within the formalism of \emph{symmetric monoidal categories} \cite{Selinger2011}, and more precisely \emph{PROPs}. In this section, we give a term-based definition of this structure and present the three variants of the diagrammatic equivalence problem that we consider in the rest of the paper.

\subsection{Circuits and Their Graphical Representation}

Given a set of generators $\Sigma$ equipped with two functions $\textup{dom},\textup{cod}:\Sigma\to\N$, we define circuits as terms of type $\cat{Circ}_\Sigma(n,m)$ for any $n,m\in\N$, as follows.
\begin{gather*}
    \infer{g\in\cat{Circ}_\Sigma(\textup{dom}(g),\textup{cod}(g))}{g\in\Sigma} \hspace{1em}
    \infer{\textup{id}_n\in\cat{Circ}_\Sigma(n,n)}{n\in\N} \hspace{1em}
    \infer{\sigma_{n,m}\in\cat{Circ}_\Sigma(n+m,m+n)}{n,m\in\N} \\[0.5em]
    \infer{C_1\semicolon C_2\in\cat{Circ}_\Sigma(n_1,m_2)}{C_i\in\cat{Circ}_\Sigma(n_i,m_i) & m_1=n_2} \hspace{1em}
    \infer{C_1\otimes C_2\in\cat{Circ}_\Sigma(n_1+n_2,m_1+m_2)}{C_i\in\cat{Circ}_\Sigma(n_i,m_i)}
\end{gather*}

Let $\cat{Circ}_\Sigma=\cup_{n,m\in\N}\cat{Circ}_\Sigma(n,m)$ be the collection of all circuits. We may write $C:n\to m$ when $C\in\cat{Circ}_\Sigma(n,m)$. Moreover, we extend $\textup{dom}$ and $\textup{cod}$ to circuits as follows: for any $C\in\cat{Circ}_\Sigma(n,m)$, let $\textup{dom}(C)\defeq n$ be the \emph{domain} of $C$ and $\textup{cod}(C)\defeq m$ be the \emph{codomain} of $C$. Intuitively, $\textup{dom}(C)$ represents the number of inputs and $\textup{cod}(C)$ the number of outputs. For any $n\in\N$, the circuit $\textup{id}_n$ is the identity of size $n$. For any $n,m\in\N$, the circuit $\sigma_{n,m}$ swaps two registers of size $n$ and $m$ respectively. The circuit $C_1\semicolon C_2$ represents a sequential composition of $C_1$ and $C_2$. Note that this composition is only possible when $m_1=n_2$. The circuit $C_1\otimes C_2$ represents a parallel composition of $C_1$ and $C_2$. 
Note that neither operation is commutative.

\begin{figure}[b]
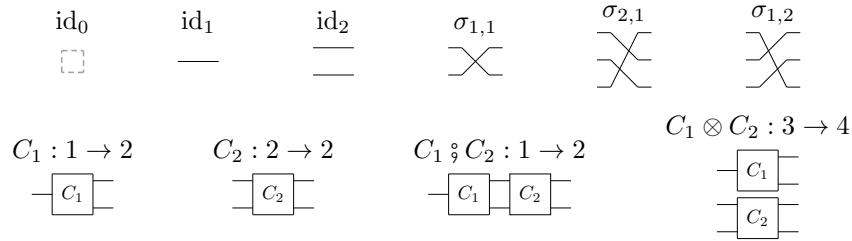

\begin{gather*}
    \begin{array}{c}
        \textup{id}_0 \\[0.2em] \smalltf{empty}
    \end{array} \hspace{2em}
    \begin{array}{c}
        \textup{id}_1 \\[0.2em] \smalltf{id1}
    \end{array} \hspace{2em}
    \begin{array}{c}
        \textup{id}_2 \\[0.2em] \smalltf{id2}
    \end{array} \hspace{2em}
    \begin{array}{c}
        \sigma_{1,1} \\[0.2em] \smalltf{sigma11}
    \end{array} \hspace{2em}
    \begin{array}{c}
        \sigma_{2,1} \\[0.2em] \smalltf{sigma21}
    \end{array} \hspace{2em}
    \begin{array}{c}
        \sigma_{1,2} \\[0.2em] \smalltf{sigma12}
    \end{array} \\[0.5em]
    \begin{array}{c}
        C_1:1\to 2 \\[0.2em] \smalltf{C1}
    \end{array} \hspace{2em}
    \begin{array}{c}
        C_2:2\to 2 \\[0.2em] \smalltf{C2}
    \end{array}\hspace{2em}
    \begin{array}{c}
        C_1\semicolon C_2:1\to2 \\[0.2em] \smalltf{seqC1C2}
    \end{array} \hspace{2em}
    \begin{array}{c}
        C_1\otimes C_2:3\to4 \\[0.2em] \smalltf{parC1C2}
    \end{array}
\end{gather*}
\caption{Examples of Circuits.}
\label{fig:prob_desc_examples}
\end{figure}

Graphically, circuits are depicted as \emph{string diagrams}, as presented in the introduction. %which consist of boxes connected by wires.
This representation makes use of the two dimensions offered by drawings.
The sequential composition is drawn horizontally and the parallel composition vertically.
In that context, a circuit with $n$ inputs and $m$ outputs is represented as a box with $n$ input wires drawn on its left-hand side and $m$ output wires drawn on its right-hand side.
There are some special cases: the circuit $\textup{id}_0$ is the empty circuit and is not represented, the circuit $\textup{id}_n$ is represented as $n$ wires in parallel when $n\ne 0$, and the circuit $\sigma_{n,m}$ is represented as a swapping of $n$ wires with $m$ wires.
We may draw dashed gray lines to identify (sub-)circuits when helpful.
Several examples are available in \cref{fig:prob_desc_examples}.

\begin{figure}
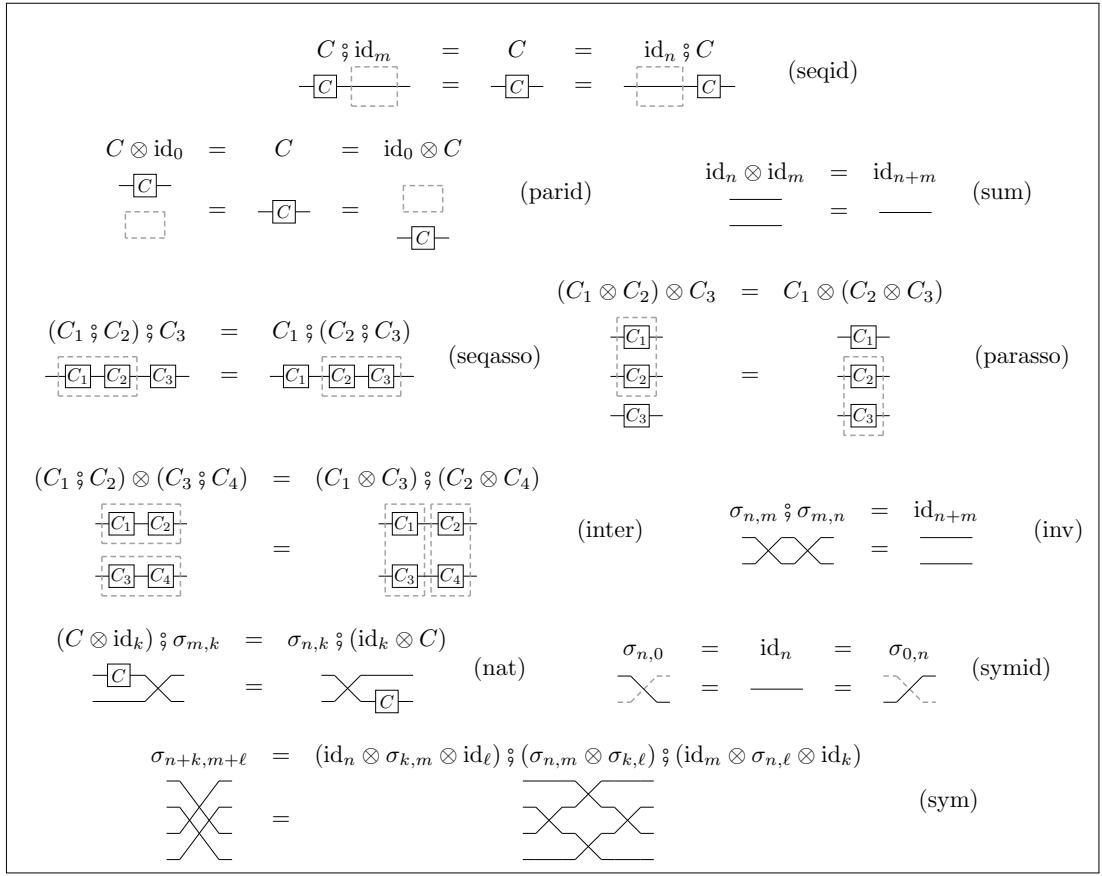

    \scalebox{0.9}{\fbox{\begin{minipage}{1.092\linewidth}\begin{center}
        \begin{subfigure}{9cm}
            \begin{equation}\label{eq:seqid}\tag{seqid}
                \begin{array}{ccccc}
                    C\semicolon\textup{id}_m&=&C&=&\textup{id}_n\semicolon C\\
                    \tf{seqidentity_right}&=&\tf{seqidentity_mid}&=&\tf{seqidentity_left}
                \end{array}
            \end{equation}
        \end{subfigure}

        \hspace{-1em}\begin{subfigure}{8.00cm}
            \begin{equation}\label{eq:parid}\tag{parid}
                \begin{array}{ccccc}
                    C\otimes \textup{id}_0&=&C&=&\textup{id}_0\otimes C\\[.4em]
                    \tf{paridentity_left}&=&\tf{paridentity_mid}&=&\tf{paridentity_right}
                \end{array}
            \end{equation}
        \end{subfigure}\hspace{3em}
        \begin{subfigure}{5.30cm}
            \begin{equation}\label{eq:sum}\tag{sum}
                \begin{array}{ccc}
                    \textup{id}_{n}\otimes\textup{id}_{m}&=&\textup{id}_{n+m}\\[.4em]
                    \tf{idinduct_left}&=&\tf{idinduct_right}
                \end{array}
            \end{equation}
        \end{subfigure}

        \hspace{-1em}\begin{subfigure}{7.80cm}
            \begin{equation}\label{eq:seqasso}\tag{seqasso}
                \begin{array}{ccc}
                    (C_1\semicolon C_2)\semicolon C_3&=&C_1\semicolon(C_2\semicolon C_3)\\[.4em]
                    \tf{seqassociativity_left}&=&\tf{seqassociativity_right}
                \end{array}
            \end{equation}
        \end{subfigure}
        \hspace{-1em}\begin{subfigure}{8cm}
            \begin{equation}\label{eq:parasso}\tag{parasso}
                \begin{array}{ccc}
                    (C_1\otimes C_2)\otimes C_3&=&C_1\otimes(C_2\otimes C_3)\\[.4em]
                    \tf{parassociativity_left}&=&\tf{parassociativity_right}
                \end{array}
            \end{equation}
        \end{subfigure}
        
        \hspace{-1em}\begin{subfigure}{9.50cm}
            \begin{equation}\label{eq:inter}\tag{inter}
                \begin{array}{ccc}
                    (C_1\semicolon C_2)\otimes(C_3\semicolon C_4)&=&(C_1\otimes C_3)\semicolon(C_2\otimes C_4)\\[.4em]
                    \tf{interchange_left}&=&\tf{interchange_right}
                \end{array}
            \end{equation}
        \end{subfigure}\hspace{1em}
        \begin{subfigure}{6.00cm}
            \begin{equation}\label{eq:inv}\tag{inv}
                \begin{array}{ccc}
                    \sigma_{n,m}\semicolon\sigma_{m,n}&=&\textup{id}_{n+m}\\[.4em]
                    \tf{involution_left}&=&\tf{involution_right}
                \end{array}
            \end{equation}
        \end{subfigure}

        \hspace{-1em}\begin{subfigure}{7.30cm}
            \begin{equation}\label{eq:nat}\tag{nat}
                \begin{array}{ccc}
                    (C\otimes\textup{id}_k)\semicolon\sigma_{m,k}&=&\sigma_{n,k}\semicolon(\textup{id}_k\otimes C)\\[.4em]
                    \tf{naturality_left}&=&\tf{naturality_right}
                \end{array}
            \end{equation}
        \end{subfigure}\hspace{2em}
        \begin{subfigure}{6.90cm}
            \begin{equation}\label{eq:symid}\tag{symid}
                \begin{array}{ccccc}
                    \sigma_{n,0}&=&\textup{id}_n&=&\sigma_{0,n}\\[.4em]
                    \tf{emptysymmetry_left}&=&\tf{emptysymmetry_mid}&=&\tf{emptysymmetry_right}
                \end{array}
            \end{equation}
        \end{subfigure}

        \hspace{-1em}\begin{subfigure}{13.00cm}
            \begin{equation}\label{eq:sym}\tag{sym}
                \begin{array}{ccc}
                    \sigma_{n+k,m+\ell}&=&(\textup{id}_n\otimes\sigma_{k,m}\otimes\textup{id}_\ell)\semicolon(\sigma_{n,m}\otimes\sigma_{k,\ell})\semicolon(\textup{id}_m\otimes\sigma_{n,\ell}\otimes\textup{id}_k)\\[.4em]
                    \tf{symmetryinduct_left}&=&\tf{symmetryinduct_right}
                \end{array}
            \end{equation}
        \end{subfigure}
        \vspace{0.2em}
    \end{center}\end{minipage}}}
    \caption{\label{fig:coherencelaws} Coherence equations defined for any $C\in\cat{Circ}_\Sigma(n,m)$ and any $C_i\in\cat{Circ}_\Sigma(n_i,m_i)$ where \cref{eq:seqasso} has to be satisfied whenever $m_1=n_2$ and $m_2=n_3$, and \cref{eq:inter} has to be satisfied whenever $m_1=n_2$ and $m_3=n_4$. Notice that a single wire depicts arbitrarily many wires to simplify the drawings.}
\end{figure}

In order to precisely capture the intuitive notion of circuits, we need equivalence equations known as \emph{coherence equations} to ensure that syntactically distinct terms that are represented by the same diagram are equal. The coherence equations are defined in \cref{fig:coherencelaws}.

Let $\cat{Circ}_\Sigma\vdash \cdot=\cdot$ be the smallest congruence relation satisfying the coherence equations. That is, $\cat{Circ}_\Sigma\vdash C_1=C_2$ means that $C_1$ can be rewritten into $C_2$ by applying the coherence equations to sub-circuits.
This relation corresponds exactly to the intuitive notion of \emph{diagrammatic equivalence} between circuits. This result is known as \emph{coherence theorem} and relates equivalent circuits to isomorphic graphs \cite{Selinger2011}.

\subsection{Challenging Equivalence Checking Problems}

This well-studied formalism of circuits offers an interesting equivalence checking problem. In this paper, we propose three variants of the problem together with their encoding in \smt{} and \tptp, yielding challenging benchmarks.

The first problem, called \emph{Circuit Equivalence} (CE), is the most general and corresponds to the decision problem associated with the equivalence relation $\cat{Circ}_\Sigma\vdash \cdot=\cdot$.
\begin{problem}
    Given $C_1,C_2\in\cat{Circ}_\Sigma$, decide whether $\cat{Circ}_\Sigma\vdash C_1=C_2$.
\end{problem}

The second problem, namely \emph{Permutation Circuit Equivalence} (PCE), is a particular instance of CE in which $\Sigma=\varnothing$.
When there are no generators, circuits are solely made of wires and always have an identical number of inputs and outputs.
In fact, any circuit of size $n$ can be seen as a permutation, i.e., a bijective map from $\{1,\dots,n\}$ to itself. The PCE problem is defined as follows.
\begin{problem}
    Given $C_1,C_2\in\cat{Circ}_\varnothing$, decide whether $\cat{Circ}_\varnothing\vdash C_1=C_2$.
\end{problem}

The third problem, referred to as \emph{Simplified Permutation Circuit Equivalence} (SPCE), is a restricted version of PCE.
It can express only concrete instances of PCE rather than families of problems parameterized by an arbitrary number of wires.
More precisely, to express any permutation, we only need the two atomic circuits $\textup{id}\defeq\textup{id}_1$ and $\sigma\defeq\sigma_{1,1}$, together with the two compositions $\semicolon$ and $\otimes$. 
The collections of \emph{simplified permutation circuits} $\cat{Perm}(n)$ are defined as follows.
\begin{gather*}
    \infer{\textup{id}\in\cat{Perm}(1)}{} \hspace{2em}
    \infer{\sigma\in\cat{Perm}(2)}{} \\[0.5em]
    \infer{C_1\semicolon C_2\in\cat{Perm}(n)}{C_i\in\cat{Perm}(n)} \hspace{2em}
    \infer{C_1\otimes C_2\in\cat{Perm}(n_1+n_2)}{C_i\in\cat{Perm}(n_i)}
\end{gather*}

\begin{figure}[tbp]
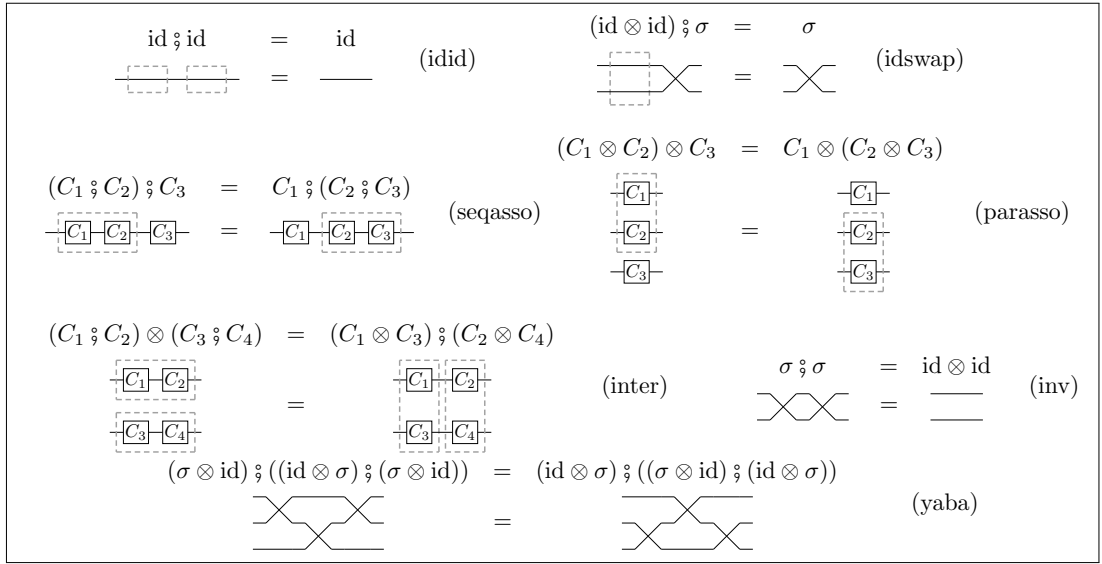

    \scalebox{0.9}{\fbox{\begin{minipage}{1.092\linewidth}\begin{center}
        \vspace{-1em}
        \hspace{-1em}\begin{subfigure}{6.00cm}
            \begin{equation}\label{eq:simp:idid}\tag{idid}
                \begin{array}{ccc}
                    \textup{id}\semicolon\textup{id}&=&\textup{id}\\[.4em]
                    \tf{idid}&=&\tf{id}
                \end{array}
            \end{equation}
        \end{subfigure}\hspace{3em}
        \begin{subfigure}{6.00cm}
            \begin{equation}\label{eq:simp:idsym}\tag{idswap}
                \begin{array}{ccc}
                    (\textup{id}\otimes\textup{id})\semicolon\sigma&=&\sigma\\[.4em]
                    \tf{idswap_left}&=&\tf{idswap_right}
                \end{array}
            \end{equation}
        \end{subfigure}\hspace{2em}

        \hspace{-1em}\begin{subfigure}{7.80cm}
            \begin{equation}\label{eq:simp:seqasso}\tag{seqasso}
                \begin{array}{ccc}
                    (C_1\semicolon C_2)\semicolon C_3&=&C_1\semicolon(C_2\semicolon C_3)\\[.4em]
                    \tf{seqassociativity_left}&=&\tf{seqassociativity_right}
                \end{array}
            \end{equation}
        \end{subfigure}\hspace{-1em}
        \begin{subfigure}{8.00cm}
            \begin{equation}\label{eq:simp:parasso}\tag{parasso}
                \begin{array}{ccc}
                    (C_1\otimes C_2)\otimes C_3&=&C_1\otimes(C_2\otimes C_3)\\[.4em]
                    \tf{parassociativity_left}&=&\tf{parassociativity_right}
                \end{array}
            \end{equation}
        \end{subfigure}
        
        \hspace{-1em}\begin{subfigure}{9.8cm}
            \begin{equation}\label{eq:simp:inter}\tag{inter}
                \begin{array}{ccc}
                    (C_1\semicolon C_2)\otimes(C_3\semicolon C_4)&=&(C_1\otimes C_3)\semicolon(C_2\otimes C_4)\\[.4em]
                    \tf{interchange_left}&=&\tf{interchange_right}
                \end{array}
            \end{equation}
        \end{subfigure}\hspace{1.5em}
        \begin{subfigure}{5.4cm}
            \begin{equation}\label{eq:simp:inv}\tag{inv}
                \begin{array}{ccc}
                    \sigma\semicolon\sigma&=&\textup{id}\otimes\textup{id}\\[.4em]
                    \tf{involution_left}&=&\tf{involution_right}
                \end{array}
            \end{equation}
        \end{subfigure}

        \hspace{-1em}\begin{subfigure}{13.00cm}
            \begin{equation}\label{eq:simp:yaba}\tag{yaba}
                \begin{array}{ccc}
                    (\sigma\otimes\textup{id})\semicolon((\textup{id}\otimes\sigma)\semicolon (\sigma\otimes\textup{id}))&=&(\textup{id}\otimes\sigma)\semicolon((\sigma\otimes\textup{id})\semicolon(\textup{id}\otimes\sigma))\\[.4em]
                    \tf{yangbaxter_left}&=&\tf{yangbaxter_right}
                \end{array}
            \end{equation}
        \end{subfigure}
        \vspace{0.2em}
    \end{center}\end{minipage}}}
    \caption{\label{fig:simpcoherencelaws} Simplified coherence equations defined for any $C_i\in\cat{Perm}(n_i)$ where \cref{eq:simp:seqasso} has to be satisfied whenever $n_1=n_2=n_3$, and \cref{eq:simp:inter} has to be satisfied whenever $n_1=n_2$ and $n_3=n_4$.}
\end{figure}

Let $\cat{Perm}=\cup_{n\in\N}\cat{Perm}(n)$ be the collection of all simplified permutation circuits. 
In this representation, the coherence equations can also be simplified and are given in \cref{fig:simpcoherencelaws}.
This new representation comes with a few noteworthy changes:
\begin{itemize}
    \item The rules \eqref{eq:simp:seqasso}, \eqref{eq:simp:parasso} and \eqref{eq:simp:inter} are unchanged.
    \item The \eqref{eq:parid} rules have disappeared, since it is no longer possible to represent the empty circuit.
    \item The \eqref{eq:seqid} rules need to be defined only for $\textup{id}$ and $\sigma$, becoming \eqref{eq:simp:idid} and \eqref{eq:simp:idsym}. 
    Notice that there is exactly one equation in \eqref{eq:simp:idid} and one equation in \eqref{eq:simp:idsym} rather than two in each.
    In the case of \eqref{eq:simp:idid}, this is because the two equations in \eqref{eq:seqid} coincide when $t=\id$.
    For \eqref{eq:simp:idsym}, one side (e.g., $\sigma\semicolon(\textup{id}\otimes\textup{id})$) can be derived from the other (e.g., $(\textup{id}\otimes\textup{id})\semicolon\sigma$) using \eqref{eq:simp:inv} and \eqref{eq:simp:seqasso}. %\seq(\swap,\stack(\id,\id)) \seq(\stack(\id,\id),\swap)$
    \item The \eqref{eq:nat} rule is replaced by a slightly simpler rule called \eqref{eq:simp:yaba} after its proposers, Yang and Baxter \cite{JoyalStreet1993}.
    Note that, due to the associativity of $\semicolon$ and $\otimes$, this rule can be represented in several equivalent forms as an equation on terms.
    \item The \eqref{eq:sum}, \eqref{eq:sym}, and \eqref{eq:symid} rules have disappeared, as identities and swaps now have fixed domains and codomains. 
\end{itemize}

The SPCE problem is defined as follows. 
\begin{problem}
    Given $C_1,C_2\in\cat{Perm}$, decide whether $\cat{Perm}\vdash C_1=C_2$.
\end{problem}

%\textit{Remark.}
\begin{remark}
To our knowledge, the complexity of the circuit equivalence problem has not been established in general. 
However, in the CE case, the problem can be embedded into graph isomorphism, placing it in the quasi-polynomial time class~\cite{babai2016graph}.
\end{remark}

\section{Encodings}
\label{sec:encodings}
We present three encodings in first-order logic with equality for the problems introduced in the previous section, using linear arithmetic to represent constraints on rule application.

\subsection{Encoding of Circuit Equivalence}

Let $\Sigma$ be a fixed set of symbols representing generators.
We introduce the following syntax to encode the terms of the type system $\cat{Circ}_\Sigma$ as first-order terms:
\[
t ::= \id(n) \mid \swap(n,m) \mid \gen(g,n,m) \mid \seq(t,t) \mid \stack(t,t)\text{ for }n,m\in\mathbb{N},g\in\Sigma.
\]

The interpretation is straightforward: $\id(n)$ is $\textup{id}_n$ and $\swap(n,m)$ is $\sigma_{n,m}$; the term $\gen(g,n,m)$ represents the generator $g$ with $n$ input wires and $m$ output wires; circuits represented by terms $t_1$ and $t_2$ are put in sequence using $\seq(t_1,t_2)$ and in parallel using $\stack(t_1,t_2)$.

In the encoding, the functions $\win$ and $\wout$ are computed as follows.
\begin{gather*}
    \begin{array}{rcl}
        \win(\id(n)) &=& n \\
        \win(\swap(n,m)) &=& n+m \\
        \win(\gen(g,n,m)) &=& n \\
        \win(\seq(t_1,t_2)) &=& \win(t_1) \\
        \win(\stack(t_1,t_2)) &=& \win(t_1)+\win(t_2)
    \end{array}
    \hspace{1em}
    \begin{array}{rcl}
        \wout(\id(n)) &=& n \\
        \wout(\swap(n,m)) &=& m+n \\
        \wout(\gen(g,n,m)) &=& m \\
        \wout(\seq(t_1,t_2)) &=& \wout(t_2) \\
        \wout(\stack(t_1,t_2)) &=& \wout(t_1)+\wout(t_2)
    \end{array}
\end{gather*}

The main difference between this syntax and that of circuits is that we define $\win(g)$ and $\wout(g)$ directly from terms instead of considering these functions already defined on generators.
Thus, two generators with the same name but different numbers of in- or outgoing wires will be considered distinct.

In fact, terms are an over-approximation of circuits.
Consider, for example, the term $\seq(\id(2),\id(3))$.
It is syntactically correct but does not correspond to any well-typed circuit in $\cat{Circ}_\Sigma(2,3)$ since $\wout(\id(2)) = \textup{cod}(\textup{id}_2)=2\neq 3= \textup{dom}(\textup{id}_3)=\win(\id(3))$, due to the mismatch in the number of wires in $\textup{id}_2$ and $\textup{id}_3$.
We call \emph{proper} the terms that correspond to circuits.
Using $\win$ and $\wout$, we can define proper terms by induction: $\id(n)$, $\swap(n,m)$ and $\gen(g,n,m)$ are proper for all $n,m\in\mathbb{N}$, $\stack(t_1,t_2)$ is proper if $t_1$ and $t_2$ are proper, and $\seq(t_1,t_2)$ is proper if $t_1$ and $t_2$ are proper and if $\wout(t_1)=\win(t_2)$.
In that way, proper terms correspond exactly to circuits.

The coherence equations form an equivalence relation on circuits, and thus on proper terms.
In our encoding, described in \cref{fig:enccoherencelaws}, we use this relation as the equality relation between terms, but there is a catch.
Some equations hold unconditionally in one direction, and with a condition in the other direction.
By using them unguarded, one can equate proper terms with improper terms, leading to contradictions.
\usetagform{Enc1}
\begin{example}
    Let $t$ be a term such that $\win(t)=2$.
    Using \eqref{eq:seqidentity} without guard, we can obtain $\seq(\id(3),t)=t$, and thus, $2=\win(t)=\win(\seq(\id(3),t))=3$, which is a contradiction.
\end{example}
We solve this issue by adding guards to ambiguous equations, ensuring that proper terms can only be equal to proper terms.
For \eqref{eq:seqidentity}, it is enough to replace the equations $\seq(\id(n),t) = t$ and $\seq(t,\id(n))=t$ by the implications:
\begin{align*}
\win(t)=n &\longrightarrow \seq(\id(n),t)=t,\\
\wout(t)=n &\longrightarrow \seq(t,\id(n))=t.
\end{align*}

The only other equation that needs a guard is \eqref{eq:interchange}, which would be as follows without guard % be naively expressed as
\begin{align*}
\stack(\seq(t,u),\seq(v,w)) &= \seq(\stack(t,v),\stack(u,w)). 
\end{align*}
Indeed, having a proper term on the right-hand side of interchange does not guarantee that the left-hand side is also proper (whereas a proper left-hand side ensures a proper right-hand side).
To preserve properness from right to left, the term has to additionally verify $\wout(t)=\win(u)$ and $\wout(v)=\win(w)$.
However, the latter can be omitted from the encoding because it follows from the former and from $\seq(\stack(t,v),\stack(u,w))$ being proper: properness implies $\wout(\stack(t,v))=\win(\stack(u,w))$, i.e.\ $\wout(t)+\wout(v)=\win(u)+\win(w)$, which together with the guard $\wout(t)=\win(u)$ yields $\wout(v)=\win(w)$.
The other equations are encoded in a straightforward way.

\usetagform{Enc1}
\begin{figure}[tbp]
    \fbox{\begin{minipage}{0.982\linewidth}\begin{center}
        %\vspace{-1em}
        \begin{subfigure}{0.55\textwidth}
            \begin{equation}\label{eq:seqidentity}\tag{seqid}
                \begin{array}{ccc}
                    \win(t) = n &\longrightarrow& \seq(\id(n),t) = t\\
                    \wout(t) = n &\longrightarrow& \seq(t,\id(n)) = t
                \end{array}
            \end{equation}
        \end{subfigure}
        \hspace{0em}
        \begin{subfigure}{0.35\textwidth}
            \begin{equation}\label{eq:paridentity}\tag{parid}
                \begin{array}{ccc}
                    \stack(t,\id(0)) &=& t\\
                    \stack(\id(0),t) &=& t
                \end{array}
            \end{equation}
        \end{subfigure}

        \begin{subfigure}{0.7\textwidth}
            \begin{equation}\label{eq:idinduct}\tag{sum}
                \begin{array}{ccc}
                    \id(n+m)&=&\stack(\id(n),\id(m))
                \end{array}
            \end{equation}
        \end{subfigure}

        \begin{subfigure}{0.7\textwidth}
            \begin{equation}\label{eq:seqassociativity}\tag{seqasso}
                \begin{array}{ccc}
                    \seq(\seq(t,u),v) &=& \seq(t,\seq(u,v))
                \end{array}
            \end{equation}
        \end{subfigure}
        \begin{subfigure}{.7\textwidth}
            \begin{equation}\label{eq:parassociativity}\tag{parasso}
                \begin{array}{ccc}
                    \stack(\stack(t,u),v) &=& \stack(t,\stack(u,v))
                \end{array}
            \end{equation}
        \end{subfigure}
        \begin{subfigure}{\textwidth}
            \begin{equation}\label{eq:interchange}\tag{inter}
                \begin{array}{c}
                    \wout(t)=\win(u)\longrightarrow
                    \stack(\seq(t,u),\seq(v,w))=\seq(\stack(t,v),\stack(u,w))
                \end{array}
            \end{equation}
        \end{subfigure}

        \begin{subfigure}{0.45\textwidth}
            \begin{equation}\label{eq:emptysymmetry}\tag{symid}
                \begin{array}{ccc}
                    \swap(n,0)&=&\id(n)\\
                    \swap(0,n)&=&\id(n)
                \end{array}
            \end{equation}
        \end{subfigure}

        \begin{subfigure}{0.7\textwidth}
            \begin{equation}\label{eq:involution}\tag{inv}
                \begin{array}{ccc}
                    \seq(\swap(n,m),\swap(m,n)) = \id(n+m)
                \end{array}
            \end{equation}
        \end{subfigure}
        \begin{subfigure}{\textwidth}
            \begin{equation}\label{eq:naturality}\tag{nat}
                \begin{array}{ccc}
                    \seq(\stack(u,\id(k)),\swap(\wout(u),k)) &=& \seq(\swap(\win(u),k),\stack(\id(k),u))\\
                \end{array}
            \end{equation}
        \end{subfigure}
        \begin{subfigure}{0.88\textwidth}
            \begin{equation}\label{eq:symmetryinduct}\tag{sym}
                \begin{array}{ccl}
                    \swap(n+k,m+l) &=& \seq(\seq(\stack(\id(n),\stack(\swap(k,m),\id(l))),\\
                    && \quad\quad\quad\quad\stack(\swap(n,m),\swap(k,l))),\\
                    && \quad\quad\stack(\id(m),\stack(\swap(n,l),\id(k))))\\
                \end{array}
            \end{equation}
        \end{subfigure}
        \vspace{0.2em}
    \end{center}\end{minipage}}
    \caption{\label{fig:enccoherencelaws} Encoding of the coherence equations of circuits.}
\end{figure}

\subsection{Encoding of Permutation Circuit Equivalence}

Next, we present two encodings, one for the PCE problem and one for the SPCE problem.

\subsubsection{Straightforward Encoding of Permutation Circuits.}
For permutation circuits, $\Sigma = \emptyset$, so the previous syntax can be simplified by removing $\gen$ altogether as follows:
\[
t ::= \id(n) \mid \swap(n,m) \mid \seq(t,t) \mid \stack(t,t)\text{ for }n,m\in\mathbb{N}.
\]
This simplification does not affect the equations of \cref{fig:enccoherencelaws}.

\subsubsection{Simplified Encoding of Permutation Circuits.}
To make this problem more accessible to theorem provers, we also designed an encoding of SPCE to reduce the amount of arithmetic reasoning required. The syntax is affected as follows:
\[
t ::= \id \mid \swap \mid \seq(t,t) \mid \stack(t,t).
\]
The $\win$ and $\wout$ functions are now such that $\win(\id)=\wout(\id)=1$ and $\win(\swap)=\wout(\swap)=2$.
They are unchanged on $\seq$ and $\stack$.
\usetagform{Enc2}
\begin{figure}[tbp]
    \fbox{\begin{minipage}{0.982\linewidth}\begin{center}
        \vspace{-1em}
        \hspace{-0.8em}
        \begin{subfigure}{0.36\textwidth}
            \begin{equation}\label{eq:idid}\tag{idid}
                \begin{array}{ccc}
                    \seq(\id,\id)=\id
                \end{array}
            \end{equation}
        \end{subfigure}\hspace{2em}
        \begin{subfigure}{0.50\textwidth}
            \begin{equation}\label{eq:idswap}\tag{idswap}
                \begin{array}{ccc}
                    \seq(\stack(\id,\id),\swap)&=&\swap
                \end{array}
            \end{equation}
        \end{subfigure}
        \begin{subfigure}{0.7\textwidth}
            \begin{equation}\label{eq:simpseqassociativity}\tag{seqasso}
                \begin{array}{ccc}
                    \seq(\seq(t,u),v)&=&\seq(t,\seq(u,v))
                \end{array}
            \end{equation}
        \end{subfigure}\hspace{0em}
        \begin{subfigure}{0.7\textwidth}
            \begin{equation}\label{eq:simpparassociativity}\tag{parasso}
                \begin{array}{ccc}
                    \stack(\stack(t,u),v)&=&\stack(t,\stack(u,v))
                \end{array}
            \end{equation}
        \end{subfigure}
        \begin{subfigure}{\textwidth}
            \begin{equation}\label{eq:simpinterchange}\tag{inter}
                \begin{array}{ccc}
                    \win(t)=\win(u)&\longrightarrow&
                    \stack(\seq(t,u),\seq(v,w))=\seq(\stack(t,v),\stack(u,w))
                \end{array}
            \end{equation}
        \end{subfigure}
        \begin{subfigure}{0.5\textwidth}
            \begin{equation}\label{eq:simpinvolution}\tag{inv}
                \begin{array}{ccc}
                    \seq(\swap,\swap)&=&\stack(\id,\id)
                \end{array}
            \end{equation}
        \end{subfigure}\hspace{2em}
        \begin{subfigure}{0.80\textwidth}
            \begin{equation}\label{eq:simpyangbaxter}\tag{yaba}
                \begin{array}{c}
                    \seq(\stack(\swap,\id),\seq(\stack(\id,\swap),\stack(\swap,\id))) = \\
                    \quad\quad\seq(\stack(\id,\swap),\seq(\stack(\swap,\id),\stack(\id,\swap)))
                \end{array}
            \end{equation}
        \end{subfigure}
        \vspace{0.2em}
    \end{center}\end{minipage}}
    \caption{\label{fig:permcoherence} Encoding of the Coherence equations of Simplified Permutation Circuits.}
\end{figure}
This new syntax gives rise to the simplified equations described in \cref{fig:permcoherence}, where the changes in the encoding mirror those from PCE to SPCE (i.e., from \cref{fig:coherencelaws} to \cref{fig:simpcoherencelaws}).

\begin{remark}Unfortunately, we are not completely rid of arithmetic constraints, since the guard from
\usetagform{Enc1}
        \eqref{eq:interchange} is still used.
        We attempted to exploit term-ordering constraints to ensure that the 
\usetagform{Enc2}
        \eqref{eq:simpinterchange} rule is always oriented left-to-right, removing the need for the guard.
        However, we determined that it was not possible to preserve this orientation over all equivalent terms, so this idea had to be abandoned. 
\end{remark}

\subsection{Encoding Files}

For each problem variant, we provide theory files encoding the corresponding coherence equations in \smt~\cite{BarST-SMT-10} and \tptp~\cite{Sut22} formats~\cite{zenodo_scripts}. Each file formalizes the corresponding syntax of circuit terms and the coherence equations described in \cref{sec:encodings}.
Benchmark instances are produced as satisfiability problems by asserting the negation of an equality between two generated terms.

\paragraph{SMT-LIB.}
We provide the \smt{}\textsf{v2} theory files:
\texttt{encoding\_CE.smt2}, \texttt{encoding\_PCE.smt2}, and \texttt{encoding\_SPCE.smt2}.
All three files make use of the \textsf{UFLIA} logic, combining uninterpreted functions with linear 
integer arithmetic, the latter being required to express the guards on the
(\ref{eq:seqid}) and (\ref{eq:inter}) rules.

\paragraph{TPTP.}
We also provide encoding files in \tptp{}: \texttt{encoding\_CE.ax}, \texttt{encoding\_PCE.ax}, and \texttt{encoding\_SPCE.ax}. All three files make use 
of the \texttt{TFF} format, which supports typed first-order logic with arithmetic, again required 
to handle the guards on rules.

\section{Benchmark Generators}
\label{sec:generators}
In addition to the encoding files, we provide a collection of benchmark instances together with the tools used to generate them.
The benchmarks target the three circuit equivalence problems: CE, PCE, and SPCE. 
For each problem class, we provide dedicated \texttt{Python} scripts that automatically generate equality problems between two syntactically distinct yet diagrammatically equivalent circuit terms. All generators follow a two-phase approach: first producing equivalent string diagrams, then translating them into terms.
Note that a single diagram can have multiple encodings. Those files are available in Zenodo~\cite{zenodo_scripts}. 

\subsection{Generation of SPCE Instances}

\usetagform{Enc2}
The script \texttt{main\_SPCE.py} generates benchmarks for the SPCE problem. The benchmark generator operates on a graphical ASCII representation of circuits, which is first produced from the script parameters and then transformed into an equivalent one by applying rewrite rules.

\begin{figure}
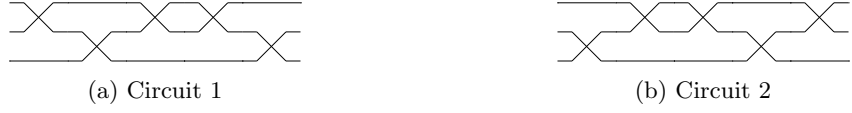

    \centering
    \begin{subfigure}[t]{0.5\linewidth}
            \centering
    \tf{spce_1}
            \caption{Circuit 1}
\label{fig:spce_c1}
    \end{subfigure}%
        \begin{subfigure}[t]{0.5\linewidth}
            \centering
    \tf{spce_2}
            \caption{Circuit 2}
\label{fig:spce_c2}
    \end{subfigure}%
    \caption{Two Equivalent Circuits for the SPCE problem.}
    \label{fig:circ_spce}
\end{figure}

\begin{figure}
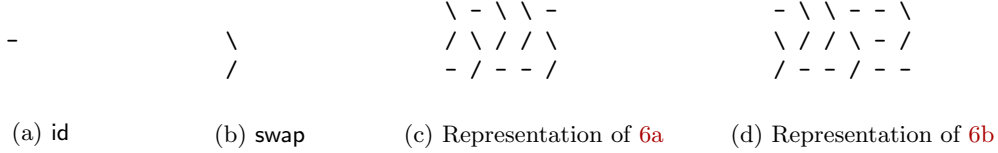

    \centering
        \begin{subfigure}[t]{0.2\linewidth}
\begin{verbatim}

      -

\end{verbatim} 
\caption{$\id$}
    \end{subfigure}%
    \begin{subfigure}[t]{0.2\linewidth}
\begin{verbatim}

      \
      /
\end{verbatim}%
\caption{$\swap$}
    \end{subfigure}%
    \begin{subfigure}[t]{0.3\linewidth}
\begin{verbatim}
      \ - \ \ - 
      / \ / / \ 
      - / - - / 
\end{verbatim} 
\caption{Representation of \ref{fig:spce_c1}}
\label{fig:spce_c1_script}
    \end{subfigure}%
    \begin{subfigure}[t]{0.3\linewidth}
\begin{verbatim}
      - \ \ - - \ 
      \ / / \ - / 
      / - - / - - 
\end{verbatim} 
\caption{Representation of \ref{fig:spce_c2}}
\label{fig:spce_c2_script}
    \end{subfigure}
\caption{ASCII Representation of Terms for the SPCE Problem.}
\label{fig:spce_script}
\end{figure}

In this setting, circuits are built exclusively from the generators $\id$ (represented by \texttt{-}) and $\swap$ (represented by {\verb $\$} for the descending wire and \texttt{/} for the ascending wire), together with sequential and parallel compositions. Example SPCE circuits are presented in \cref{fig:circ_spce}, while their corresponding ASCII representations, as well as additional terms, are available in \cref{fig:spce_script}.

Internally, circuits are represented as rectangular grids, where rows correspond to wires and columns correspond to successive layers of sequential compositions.
The script is parameterized by the number of rows (\texttt{-r}) and columns (\texttt{-c}), which directly controls the size of the original circuit. The \texttt{-s} flag gives the number of rule applications. 
For instance, circuits \ref{fig:spce_c1_script} and \ref{fig:spce_c2_script} were generated by \texttt{main\_SPCE.py -r=3 -c=5 -s 100}. 

Starting from an initial number of rows $r$ and columns $c$, the benchmark generator starts with an $\id$-only grid of size $r \times c$ and then randomly adds binary swaps. 
Then, in order to produce an equivalent circuit, the benchmark generator selects a rectangular subpart of the grid and applies a sequence of transformations corresponding to a subset of the simplified coherence equations \eqref{eq:idid}, \eqref{eq:idswap}, \eqref{eq:simpinvolution}, and \eqref{eq:simpyangbaxter} of \cref{fig:permcoherence}. Note that the \eqref{eq:idid} rule can modify the number of columns of the circuit.

The remaining rules are handled by the translation into terms. In practice, the benchmark generator selects a split direction (vertical or horizontal), then checks if the cut is feasible (\ie{} by not breaking a $\swap$). If the split is allowed, a sequential ($\seq$) or parallel ($\stack$) composition of the two sub-circuits thus obtained is generated.

\subsection{Generation of PCE Instances}
\begin{figure}
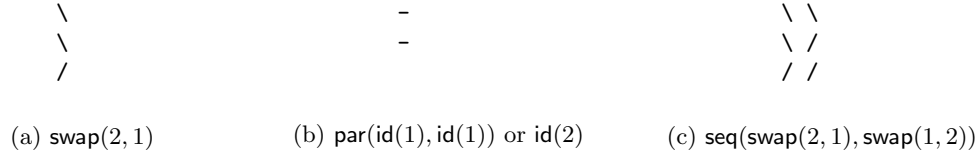

    \centering
    \begin{subfigure}[t]{0.3\linewidth}
\begin{verbatim}
           \
           \
           /
\end{verbatim} 
\caption{$\swap(2,1)$}
    \end{subfigure}%
    \begin{subfigure}[t]{0.35\linewidth}
\begin{verbatim}
            -
            -

\end{verbatim} 
\caption{$\stack(\id(1), \id(1))$ or $\id(2)$}
    \end{subfigure}%
    \begin{subfigure}[t]{0.35\linewidth}
\begin{verbatim}
            \ \ 
            \ / 
            / /
\end{verbatim} 
\caption{$\seq(\swap(2,1),\swap(1, 2))$}
    \end{subfigure}
\caption{Representation of Elements of PCE.}
\label{fig:pce_script}
\end{figure}

\usetagform{Enc1}
The PCE script \texttt{main\_PCE.py} extends the previous script to handle permutation circuits with arbitrary-size swaps, as illustrated in \cref{fig:pce_script}. It takes an additional parameter \texttt{-a} that controls the maximal size of a swap.

Unlike the previous script, this one starts by randomly inserting $n$-ary swaps, together with two other patterns: \eqref{eq:symmetryinduct} and \eqref{eq:involution}, of which the ASCII representation is available in \cref{fig:pce_patterns}. 
These patterns can randomly arise with the insertion of swaps.
However, to produce more realistic and diverse circuits, we chose to include dedicated functions to add them.

\begin{figure}
\centering
\begin{minipage}{0.48\linewidth}
\centering

\begin{subfigure}[t]{0.45\linewidth}
\centering
\begin{verbatim}
        \ \ 
        \ / 
        / / 
\end{verbatim}
\caption{Left Pattern}
\end{subfigure}\hfill
\begin{subfigure}[t]{0.45\linewidth}
\centering
\begin{verbatim}
        - -
        - -
        - -
\end{verbatim}
\caption{Right Pattern}
\end{subfigure}
\caption*{\eqref{eq:involution} Rule Patterns for $\swap(2,1)$}
\end{minipage}
\hfill
\begin{minipage}{0.48\linewidth}
\centering

\begin{subfigure}[t]{0.45\linewidth}
\centering
\begin{verbatim}
        \   
        \   
        /   
        /   
\end{verbatim}
\caption{Left Pattern}
\end{subfigure}\hfill
\begin{subfigure}[t]{0.45\linewidth}
\centering
\begin{verbatim}
        - \ -
        \ / \
        / \ /
        - / -
\end{verbatim}
\caption{Right Pattern}
\end{subfigure}
\caption*{\eqref{eq:symmetryinduct} Rule Pattern for $\swap(2,2)$}
\end{minipage}
\caption{Two Instances of the \eqref{eq:symmetryinduct} and \eqref{eq:involution} Rules of PCE.}
\label{fig:pce_patterns}
\end{figure}

Next, as in the SPCE case, random equivalence-preserving rewrite rules are applied to generate an equivalent circuit. Three rules of \cref{fig:enccoherencelaws} are explicitly managed: sequential identity \eqref{eq:seqidentity}, involution \eqref{eq:involution}, and symmetry decomposition \eqref{eq:symmetryinduct}. The first one inserts or deletes columns of $\id$, modifying the number of columns of the grid. The latter two detect specific graphical patterns and rewrite them according to the corresponding rule. 

\begin{figure}
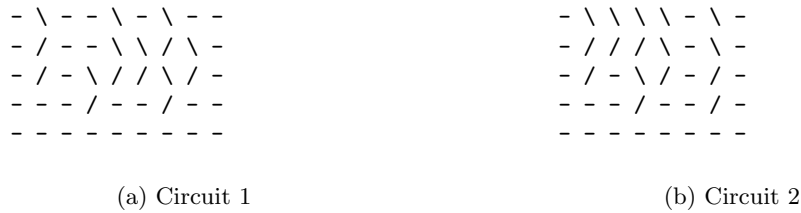

    \centering
    \begin{subfigure}[t]{0.5\linewidth}
\begin{verbatim}
        - \ - - \ - \ - -
        - / - - \ \ / \ -
        - / - \ / / \ / -
        - - - / - - / - -
        - - - - - - - - -
\end{verbatim} 
\caption{Circuit 1}
    \end{subfigure}%
    \begin{subfigure}[t]{0.5\linewidth}
\begin{verbatim}
        - \ \ \ \ - \ -
        - / / / \ - \ -
        - / - \ / - / -
        - - - / - - / -
        - - - - - - - -
\end{verbatim} 
\caption{Circuit 2}
    \end{subfigure}
\caption{Example Equivalent Circuits for PCE.}
\label{fig:pce_circuits}
\end{figure}

In practice, the benchmark generator selects a rule (identity insertion or deletion, \eqref{eq:symmetryinduct} or \eqref{eq:involution} with a direction) and retrieves all applicable positions of the rule, then chooses one and applies it to the circuit. As an example, \texttt{main\_PCE.py -r=5 -c=5 -a=4 -s=200} was used to generate the two equivalent circuits of \cref{fig:pce_circuits}. 

Moreover, an additional rule is added for $\id$ management: whenever a parallel composition of two $\id$ nodes is generated, the benchmark generator tries to rewrite it as another equivalent decomposition. For instance, $\stack(\id(2), \id(3))$ can be turned into $\id(5)$ or $\stack(\id(1), \id(4))$, ensuring more randomness. The translation into terms is analogous to the SPCE case. 
Finally, the script also includes a validation step that ensures that every input wire exits at the same position in both circuits.

\subsection{Generation of CE Instances}

The CE script (\texttt{main\_CE.py}) produces the most general benchmarks, allowing arbitrary generators. Unlike the previous benchmark generators, it is not ASCII-based; instead, it relies on an internal graph structure.

Such a graph consists of nodes, which are either identities ($\id(n)$) or generators ($\gen(n, m)$), both parameterized by their respective arities. 
Nodes maintain lists of predecessors and successors, as well as two lists corresponding to their input and output wires (including their positions and connected nodes). Finally, each node has a width, which corresponds to the number of columns it occupies.

To build a circuit, the script begins by introducing a set of generators, which are then combined into a directed planar acyclic graph. This graph is first organized into vertical layers using a topological sort, ensuring global wire connection consistency from inputs to outputs. Note that acyclicity and planarity are guaranteed by construction. Identity nodes are subsequently inserted as needed to fill the gaps and preserve wire alignment between layers. Two special nodes, \texttt{begin} and \texttt{end}, are added to collect wires that have no input or no output, respectively. Note that, in the scripts, generators with no input or no output are not allowed.

Moreover, in order to be closer to real-life circuits, we add three pre-defined generators, corresponding to Clifford generators~\cite{gottesman1998theory}: the Hadamard generator \textit{H} ($\gen(1, 1)$), the phase generator \textit{S} ($\gen(1, 1)$), and the \textit{CNOT} generator ($\gen(2, 2)$). 
A dedicated option \texttt{--clifford} forces the script to only use those predefined generators. 

This process generates the original circuit in its most compact form, where each node is assigned a row and a temporary column. Next, identity nodes are randomly inserted between existing nodes (corresponding to rule~\eqref{eq:seqidentity} in \cref{fig:enccoherencelaws}), and the column assignments of the nodes are updated accordingly.

\begin{figure}
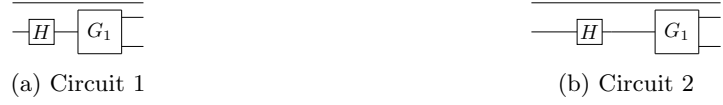

    \centering
    \begin{subfigure}[t]{0.5\linewidth}
            \centering
    \tf{ce_1}
            \caption{Circuit 1}
\label{fig:ce_c1}
    \end{subfigure}%
        \begin{subfigure}[t]{0.5\linewidth}
            \centering
    \tf{ce_2}
            \caption{Circuit 2}
\label{fig:ce_c2}
    \end{subfigure}%
    \caption{Two Equivalent Circuits \ref{fig:ce_c1} and \ref{fig:ce_c2} for CE with $2$ inputs and $3$ outputs.}
    \label{fig:circ_ce}
\end{figure}

An instance of a compact graph, together with one of its extended versions, is given in \cref{fig:circ_ce}, and its internal representation in \cref{fig:ce_circuits}. 
This circuit was generated by \texttt{main\_CE.py -r=2 -c=2}. 

\begin{figure}
    {\scriptsize
    \centering
    \begin{subfigure}[t]{1\linewidth}
\begin{verbatim}
Circuit(2,2,4):
  [0] Gen(begin, 0, 2)@([0;0], 0, 1) | pred: [-] | succ: [[2] Id(1), [3] Gen(H, 1, 1)]
  [2] Id(1)@([0;0], 1, 1) | pred: [[0] Gen(begin, 0, 2)] | succ: [[4] Id(1)]
  [3] Gen(H,1,1)@([1;1], 1, 1) | pred: [[0] Gen(begin, 0, 2)] | succ: [[5] Gen(Gen1, 1, 1)]
  [4] Id(1)@([0;0], 2, 1) | pred: [[2] Id(1)] | succ: [[1] Gen(end, 3, 0)]
  [5] Gen(Gen1, 1, 2)@([1;1], 2, 1) | pred: [[3] Gen(H, 1, 1)] | succ: [[1] Gen(end, 3, 0)]
  [1] Gen(end, 3, 0)@([0;0], 3, 1) | pred: [[4] Id(1), [5] Gen(Gen1, 1, 1)] | succ: [-]
\end{verbatim} 
\caption{Compact form of the circuit}
    \end{subfigure}
    \begin{subfigure}[t]{1\linewidth}
\begin{verbatim}
Circuit(2, 2, 6):
  [0] Gen(begin, 0, 2)@([0;0], 0, 1) | pred: [-] | succ: [[2] Id(1), [6] Id(1)]
  [2] Id(1)@([0;0], 1, 1) | pred: [[0] Gen(begin, 0, 2)] | succ: [[7] Id(1)]
  [6] Id(1)@([1;1], 1, 1) | pred: [[0] Gen(begin, 0, 2)] | succ: [[3] Gen(H, 1, 1)]
  [7] Id(1)@([0;0], 2, 1) | pred: [[2] Id(1)] | succ: [[8] Id(1)]
  [3] Gen(H, 1, 1)@([1;1], 2, 1) | pred: [[6] Id(1)] | succ: [[9] Id(1)]
  [8] Id(1)@([0;0], 3, 1) | pred: [[7] Id(1)] | succ: [[4] Id(1)]
  [9] Id(1)@([1;1], 3, 1) | pred: [[3] Gen(H, 1, 1)] | succ: [[5] Gen(Gen1, 1, 1)]
  [4] Id(1)@([0;0], 4, 1) | pred: [[8] Id(1)] | succ: [[1] Gen(end, 3, 0)]
  [5] Gen(Gen1, 1, 2)@([1;1], 4, 1) | pred: [[9] Id(1)] | succ: [[1] Gen(end, 3, 0)]
  [1] Gen(end, 3, 0)@([0;0], 5, 1) | pred: [[4] Id(1), [9] Id(1)] | succ: [-]
\end{verbatim} 
\caption{Extended version of the circuit}
    \end{subfigure}
    }
\caption{Corresponding Representation for \ref{fig:ce_c1} and \ref{fig:ce_c2}.}
\label{fig:ce_circuits}
\end{figure}

In this formalism, a circuit is parameterized by its number of inputs, outputs, and columns. A generator is represented as \texttt{Gen(name, dom, cod)}, while an identity node is written as \texttt{Id(n)}. 
The remaining information in all nodes is: 
\texttt{@([row of the first input wire; row of the first output wire], column, width) | input wires list | output wires list}.

From this underlying circuit graph, two equivalent terms are extracted by repeatedly merging adjacent nodes using either sequential or parallel composition. To this end, the algorithm first selects a node and a split direction (vertical or horizontal). It then attempts to merge the node according to the chosen direction: horizontally with its predecessor or successor, or vertically with the node above or below. If the merge is possible, a new node---either $\stack$ or $\seq$, parameterized by the two chosen nodes---is created, then added to the graph, and the two merged nodes are removed. During parallel merges, the benchmark generator randomly inserts swap nodes and, as in the PCE generator, performs identity rewritings.

%In addition to \smt{} and \tptp{} outputs, the script also provides an \isabelle{}~\cite{nipkow2002isabelle} output, corresponding to a theory presented in \cite{trs}.
The resulting instances make use of the full set of coherence equations and constitute the most challenging benchmarks in our collection.

\medskip

Overall, the three scripts generate circuits equivalent by construction, enabling the generation of large families of various circuit equivalence problems.

\section{Evaluations}
\label{sec:eval}

This section describes the experimental setup used to evaluate the benchmark families introduced in the previous sections. Our objective is to assess both the difficulty of the proposed circuit equivalence problems and the practical impact of the different encodings and tool configurations. 

\paragraph{Tools.}
Beyond performance, our two main criteria when selecting tools for this work were (i) support for arithmetic reasoning, and (ii) the ability to produce proof certificates.
The latter requirement is stronger than solving an instance and outputting a proof trace, as the certificate must be independently checkable. 
This criterion stems from the broader quantum circuit verification pipeline that motivated this work~\cite{trs}, but, although proof production was tested, a full evaluation of the verification pipeline goes beyond the scope of the current evaluation.

With respect to arithmetic reasoning, SMT solvers are generally well-suited, whereas not all first-order theorem provers provide dedicated decision procedures for arithmetic. Conversely, due to their internal reasoning techniques, traditional theorem provers are often more suitable for producing detailed proofs.

We experimented with several automated reasoning tools, including \egg~\cite{zhang2023better}, \twee~\cite{smallbone2021twee}, \eprover~\cite{schulz2019e}, \vampire~\cite{bartek2025vampire}, \zthree~\cite{de2008z3}, and \cvc~\cite{DBLP:conf/tacas/BarbosaBBKLMMMN22}. 
We first excluded provers that do not support arithmetic reasoning, namely E and Twee. We then evaluated the proof-production capabilities of the remaining tools and identified two main pipelines, thus excluding egglog and Z3.

The first one relies on \cvc{}, which can generate proofs in formats such as \lsfc{}~\cite{stump2013smt}, \eunoia{}~\cite{dunne2025automatically}, and \alethe{}~\cite{Schurr_2021}.
The latter two formats can be translated and independently checked by proof assistants such as \lp{}~\cite{lambdapi,coltellacci2024reconstruction} or \isabelle{}~\cite{lachnitt2025improving}. Although not all proof steps are currently supported, the implemented subset is sufficient for our encodings.

Regarding ATPs, \vampire{} can handle arithmetic, and it can generate SMT-style proofs from FOF encodings that are externally checkable, for instance by \cvc{}~\cite{rawson2025ground}.
Moreover, by providing problems directly in CNF to \vampire{}, we avoid its internal clausification phase, thereby enabling complete proof reconstruction in \lp{}~\cite{komel2025case}.

We therefore focus our evaluation on \vampire{} version~5.0.0 and \cvc{} version~1.3.0, as both support arithmetic reasoning and produce independently checkable proof certificates.

\paragraph{Benchmarks design.}

We partition our benchmark suite into three categories corresponding to the problem classes CE, PCE, and SPCE.

For CE, we use a mixed setting in which Clifford generators are used with probability $50\%$, while the remaining generators are of the form $\gen(n,m)$, with $n,m \leq 3$, and are produced randomly.
While our experimental choices are intended to reflect real-world circuits, these parameters can be adjusted in the benchmark generators to increase randomness. 

For each (sub-)category, we generate pairs of equivalent circuit terms and combine them with the corresponding theory file from \cref{sec:encodings}, resulting in complete solver inputs. The complete dataset used for our experiments is also available in Zenodo~\cite{zenodo_benchs}.

Instances are parameterized by two structural parameters: the number of input wires and the initial number of circuit columns.
For SPCE and PCE, both parameters range over \( \{5,10,15,20\} \), while for CE the number of input wires ranges over \( \{2,3,4,5\} \).
Since transformations can arbitrarily increase or decrease the number of circuit columns during instance generation, the initial column count alone is not sufficient to characterize instance difficulty.
To obtain a more robust measure of circuit size, we therefore additionally record the number of sequential ($\seq$) and parallel ($\stack$) composition operators occurring in the generated terms.
The total number of such compositions ranges from $0$ to $200$ and is partitioned into eight intervals of equal width (i.e., $[0,25]$, $[25,50]$, \dots).

To ensure representative instances within each interval, we restrict term sizes to lie within a $\pm 5$ window around the interval midpoint.
For example, instances in the interval $[0, 25]$ contain between $7$ and $17$ composition operators. For each interval, $100$ problems were generated.

\paragraph{Experimental setup.}

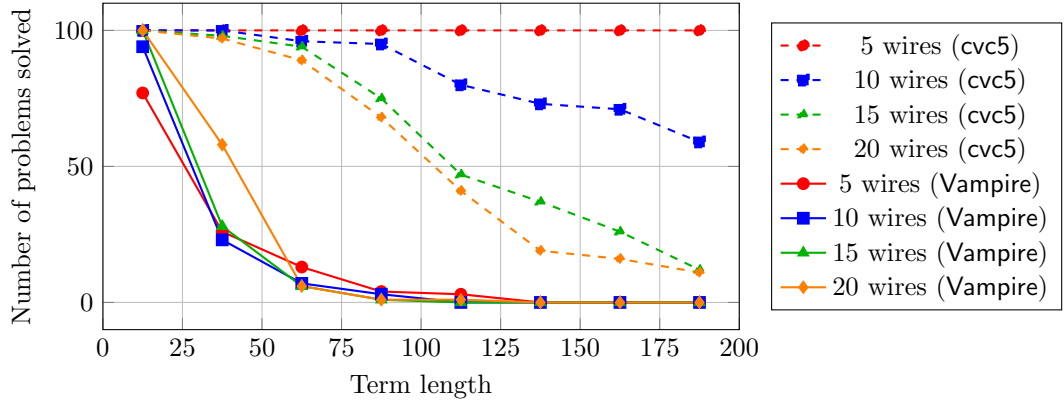
\begin{figure}
        \begin{tikzpicture}
            \begin{axis}[
                    width=10cm,
                    height=5.9cm,
                    xlabel={Term length},
                    ylabel={Number of problems solved},
                    xmin=0, xmax=200,
                    xtick={0,25,...,200},
                    grid=both,
                    legend style={
                            at={(1.05,0.5)},
                            anchor=west
                        },
                ]

                % CVC
                \addplot[
                    color=red, mark=*, thick, dashed
                ] table[
                        x=middle, y=num_theorem, col sep=comma,
                    ] {./stats/smt_SPCE_none_row_5_summary.csv};
                \addlegendentry{5 wires (\cvc)}

                \addplot[
                    color=blue, mark=square*, thick, dashed
                ] table[
                        x=middle, y=num_theorem, col sep=comma,
                    ] {./stats/smt_SPCE_none_row_10_summary.csv};
                \addlegendentry{10 wires (\cvc)}

                \addplot[
                    color=green!70!black, mark=triangle*, thick, dashed
                ] table[
                        x=middle, y=num_theorem, col sep=comma,
                    ] {./stats/smt_SPCE_none_row_15_summary.csv};
                \addlegendentry{15 wires (\cvc)}

                \addplot[
                    color=orange, mark=diamond*, thick, dashed
                ] table[
                        x=middle, y=num_theorem, col sep=comma,
                    ] {./stats/smt_SPCE_none_row_20_summary.csv};
                \addlegendentry{20 wires (\cvc)}

                % ARI
                \addplot[
                    color=red, mark=*, thick
                ] table[
                        x=middle, y=num_theorem, col sep=comma,
                    ] {./stats/tptp_SPCE_ari_row_5_summary.csv};
                \addlegendentry{5 wires (\vampire{})}

                \addplot[
                    color=blue, mark=square*, thick
                ] table[
                        x=middle, y=num_theorem, col sep=comma,
                    ] {./stats/tptp_SPCE_ari_row_10_summary.csv};
                \addlegendentry{10 wires (\vampire{})}

                \addplot[
                    color=green!70!black, mark=triangle*, thick
                ] table[
                        x=middle, y=num_theorem, col sep=comma,
                    ] {./stats/tptp_SPCE_ari_row_15_summary.csv};
                \addlegendentry{15 wires (\vampire{})}

                \addplot[
                    color=orange, mark=diamond*, thick
                ] table[
                        x=middle, y=num_theorem, col sep=comma,
                    ] {./stats/tptp_SPCE_ari_row_20_summary.csv};
                \addlegendentry{20 wires (\vampire{})}
            \end{axis}
        \end{tikzpicture}
    \caption{Results of \cvc{} and \vampire{} on SPCE Benchmarks}
    \label{fig:spce}
\end{figure}

\begin{figure}
    \begin{tikzpicture}
        \begin{axis}[
                width=10cm,
                height=5.9cm,
                xlabel={Term length},
                ylabel={Number of problems solved},
                xmin=0, xmax=200,
                xtick={0,25,...,200},
                grid=both,
                legend style={
                        at={(1.05,0.5)},
                        anchor=west
                    },
            ]

            % CVC5
            \addplot[
                color=red, mark=*, thick, dashed
            ] table[
                    x=middle, y=num_theorem, col sep=comma,
                ] {./stats/smt_CE_none_row_2_summary.csv};
            \addlegendentry{2 wires (\cvc)}

            \addplot[
                color=blue, mark=square*, thick, dashed
            ] table[
                    x=middle, y=num_theorem, col sep=comma,
                ] {./stats/smt_CE_none_row_3_summary.csv};
            \addlegendentry{3 wires (\cvc)}

            \addplot[
                color=green!70!black, mark=triangle*, thick, dashed
            ] table[
                    x=middle, y=num_theorem, col sep=comma,
                ] {./stats/smt_CE_none_row_4_summary.csv};
            \addlegendentry{4 wires (\cvc)}

            \addplot[
                color=orange, mark=diamond*, thick, dashed
            ] table[
                    x=middle, y=num_theorem, col sep=comma,
                ] {./stats/smt_CE_none_row_5_summary.csv};
            \addlegendentry{5 wires (\cvc)}

            % Vampire
             \addplot[
                color=red, mark=*, thick
            ] table[
                    x=middle, y=num_theorem, col sep=comma,
                ] {./stats/tptp_CE_none_row_2_summary.csv};
            \addlegendentry{2 wires (\vampire)}

            \addplot[
                color=blue, mark=square*, thick
            ] table[
                    x=middle, y=num_theorem, col sep=comma,
                ] {./stats/tptp_CE_none_row_3_summary.csv};
            \addlegendentry{3 wires (\vampire)}

            \addplot[
                color=green!70!black, mark=triangle*, thick
            ] table[
                    x=middle, y=num_theorem, col sep=comma,
                ] {./stats/tptp_CE_none_row_4_summary.csv};
            \addlegendentry{4 wires (\vampire)}

            \addplot[
                color=orange, mark=diamond*, thick
            ] table[
                    x=middle, y=num_theorem, col sep=comma,
                ] {./stats/tptp_CE_none_row_5_summary.csv};
            \addlegendentry{5 wires (\vampire)}

        \end{axis}
    \end{tikzpicture}
    \caption{Results of \cvc{} and \vampire{} on CE Benchmarks}
    \label{fig:ce}
\end{figure}
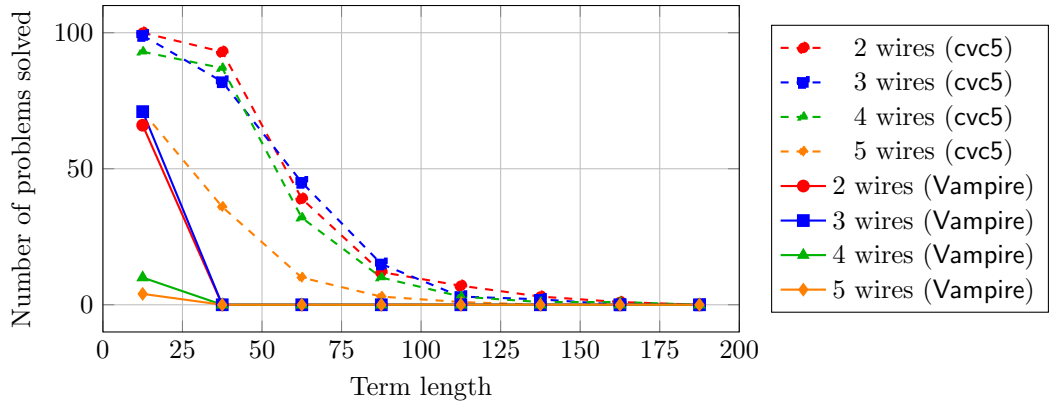

All experiments ran on the Grid5000~\cite{grid5000} cluster using an Intel Xeon Gold 5220 CPU (Cascade Lake-SP, x86\_64, 2.20\,GHz, 18 cores, 96\,GiB RAM).
For each category, we recorded the number of successfully proved equivalences.
A timeout of 30 seconds was imposed for each solver run. 
\vampire{} ran with the command line \texttt{vampire -t 30s}.
\cvc{} ran with the command line \texttt{cvc5 --tlimit 30000 --stats}.
Results are available in \cref{fig:spce,fig:pce,fig:ce}.

\begin{figure}
    \begin{tikzpicture}
        \begin{axis}[
                width=10cm,
                height=5.9cm,
                xlabel={Term length},
                ylabel={Number of problems solved},
                xmin=0, xmax=200,
                xtick={0,25,...,200},
                grid=both,
                legend style={
                        at={(1.05,0.5)},
                        anchor=west
                    },
            ]

            % CVC5
            \addplot[
                color=red, mark=*, thick, dashed
            ] table[
                    x=middle, y=num_theorem, col sep=comma,
                ] {./stats/smt_PCE_none_row_5_summary.csv};
            \addlegendentry{5 wires (\cvc)}

            \addplot[
                color=blue, mark=square*, thick, dashed
            ] table[
                    x=middle, y=num_theorem, col sep=comma,
                ] {./stats/smt_PCE_none_row_10_summary.csv};
            \addlegendentry{10 wires (\cvc)}

            \addplot[
                color=green!70!black, mark=triangle*, thick, dashed
            ] table[
                    x=middle, y=num_theorem, col sep=comma,
                ] {./stats/smt_PCE_none_row_15_summary.csv};
            \addlegendentry{15 wires (\cvc)}

            \addplot[
                color=orange, mark=diamond*, thick, dashed
            ] table[
                    x=middle, y=num_theorem, col sep=comma,
                ] {./stats/smt_PCE_none_row_20_summary.csv};
            \addlegendentry{20 wires (\cvc)}

            % Vampire
             \addplot[
                color=red, mark=*, thick
            ] table[
                    x=middle, y=num_theorem, col sep=comma,
                ] {./stats/tptp_PCE_none_row_5_summary.csv};
            \addlegendentry{5 wires (\vampire)}

            \addplot[
                color=blue, mark=square*, thick
            ] table[
                    x=middle, y=num_theorem, col sep=comma,
                ] {./stats/tptp_PCE_none_row_10_summary.csv};
            \addlegendentry{10 wires (\vampire)}

            \addplot[
                color=green!70!black, mark=triangle*, thick
            ] table[
                    x=middle, y=num_theorem, col sep=comma,
                ] {./stats/tptp_PCE_none_row_15_summary.csv};
            \addlegendentry{15 wires (\vampire)}

            \addplot[
                color=orange, mark=diamond*, thick
            ] table[
                    x=middle, y=num_theorem, col sep=comma,
                ] {./stats/tptp_PCE_none_row_20_summary.csv};
            \addlegendentry{20 wires (\vampire)}
            
        \end{axis}
    \end{tikzpicture}
    \caption{Results of \cvc{} and \vampire{} on PCE Benchmarks}
    \label{fig:pce}
\end{figure}
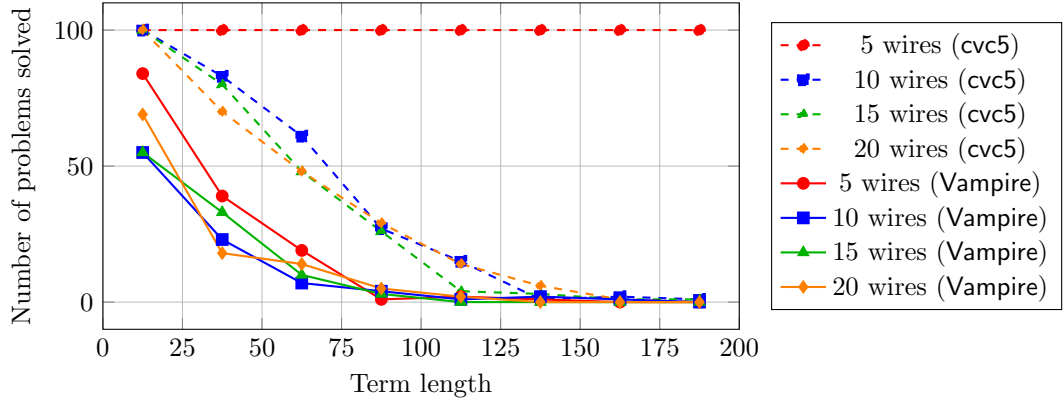

\paragraph{Results overview.}

First of all, \cvc{} outperforms \vampire{} across all benchmark families. It solves more instances overall and scales better as both the number of wires and the term size increase. In particular, for SPCE and PCE with $5$ wires, \cvc{} solves all instances across the entire range of term sizes considered. 
This is not unexpected since SMT solvers are, by design, more suitable in general to reason with arithmetic. 

Turning to structural parameters, the number of input wires and the term length have the strongest impact on the results. 
For \cvc{} on SPCE, wire count is the dominant factor: the number of solved instances drops significantly as the number of wires increases beyond $10$, regardless of term length (\cref{fig:spce}). 
By contrast, for \cvc{} on PCE and \vampire{} on SPCE and PCE, term length plays a comparably important role: performance degrades noticeably once term lengths exceed roughly $75$--$100$ composition operators (\cref{fig:spce,fig:pce}).

Finally, CE (\cref{fig:ce}) is the most challenging benchmark category. 
We restrict experiments to instances with $2$ to $5$ input wires, as handling 5 wires is already challenging.
Even within this restricted range, only small instances are solved. 
For instance, whereas \cvc{} solves all SPCE and PCE instances with $5$ wires, regardless of term length, it struggles on CE benchmarks once the term size exceeds roughly $100$. For its part, \vampire{} was not able to solve any problem above the $[0$--$25]$ interval, regardless of the number of wires.

To sum up, SPCE instances are largely solvable by \cvc{} across a wide range of parameters, PCE instances show a performance drop as both wire count and term size increase, and CE instances are the hardest, with very few problems solved beyond small configurations.

\section{Conclusion}
\label{sec:conclusion}
In this work, we presented a new family of benchmarks for diagrammatic equivalence checking, motivated by the certification of quantum circuit equivalences. We introduced three variants of the problem---CE, PCE, and SPCE---and provided first-order encodings in both \tptp{} and \smt{}, together with parameterizable scripts to generate large collections of instances.

Our experiments show that diagrammatic equivalence remains challenging for current automated provers. SMT solvers such as \cvc{} handle the guarded encodings more robustly than ATPs like \vampire{}, solving more instances across all benchmark families. Overall, the interaction between arithmetic and equational reasoning remains a central bottleneck, and both the number of input wires and the syntactic size of terms significantly impact performance, with their relative influence varying across solvers and benchmark families.

This work suggests several directions for future research. There are more ATPs with proof-production capacities than \vampire{} and \cvc{}, but most do not handle arithmetic. Thus,  an arithmetic-free encoding that correctly captures the guards would be welcome. It could also be interesting to perform further experiments varying the benchmark generation parameters (for instance, by restricting the applicable rules, or by restricting the circuits to Clifford generators only). This would enable fine-grained analyses, e.g., investigating whether some coherence rules are potential bottlenecks or if a phase transition exists for this kind of problem. This in turn could provide a principled way to generate harder benchmarks.

We hope that this benchmark suite will serve as a stress test for existing solvers and that, along with the generation scripts, they will help with the development of new techniques for reasoning about string diagrams and even quantum circuits.

\newpage
\label{sect:bib}
\bibliographystyle{splncs04}
\bibliography{bib}

\end{document}